\documentclass[aps,pre,preprint,showkeys,a4paper]{revtex4-1}
\usepackage[utf8]{inputenc}
\usepackage{booktabs}
\usepackage{graphicx,color}
\pdfoutput=1 

\begin{document}
\title{Classification of diffusion modes in single-particle tracking data: Feature-based versus deep-learning approach}
\author{Patrycja Kowalek}
\author{Hanna Loch-Olszewska}
\author{Janusz Szwabiński}
\affiliation{Faculty of Pure and Applied Mathematics, Hugo Steinhaus Center, Wrocław University of Science and Technology}

\begin{abstract}
Single-particle trajectories measured in microscopy experiments contain important information about dynamic processes undergoing in a range of materials including living cells and tissues. However, extracting that information is not a trivial task due to the stochastic nature of particles' movement and the sampling noise. In this paper, we adopt a deep-learning method known as a convolutional neural network (CNN) to classify modes of diffusion from given trajectories. We compare this fully automated approach working with raw data to classical machine learning techniques that require data preprocessing and extraction of human-engineered features from the trajectories to feed classifiers like random forest or gradient boosting.  All methods are tested  using simulated trajectories for which the underlying physical model is known. From the results it follows that
CNN is usually slightly better than the feature-based methods, but at the costs of much longer processing times. Moreover, there are still some borderline cases, in which the classical methods perform better than CNN.
\end{abstract}


\maketitle

\section{Introduction}

Recent advances in single-molecule microscopy and imaging technologies have made single-particle tracking (SPT) a popular method for analyzing dynamic processes in a range of materials~\cite{MAN15,SHE17}. In a typical SPT measurement the molecules of interest (e.g. proteins in a living cell) are tagged with fluorescent dye particles. After illumination by a laser, the labels produce light and their positions may be determined with a microscope. Using lasers that flash at short time intervals allows for tracking of the movement of the molecules over time. The recorded positions are used to reconstruct trajectories of individual molecules. These trajectories are then analyzed in order to extract local physical properties of the molecules and their environment, such as velocity, diffusion coefficient (or tensor) and confinement (local density of obstacles)~\cite{HOL15}.

The SPT method is of particular importance for fundamental biology. It bridges the gap between biology, biochemistry and biophysics and allows for at least a partial understanding of living cells on a microscopic basis. It helped already to unveil the details of the movement of molecular motors inside cells~\cite{KUR05,YIL03} or target search mechanisms of nuclear proteins~\cite{IZE14}.

The analysis of SPT trajectories is not a trivial task due to the stochastic nature of the molecules' movement.
It usually starts with the detection of a corresponding motion type of a molecule, because this information may already provide insight into mechanical properties of the molecule's surrounding~\cite{MAH09}. Four basic types of motion are observed in SPT experiments: normal diffusion (ND)~\cite{ALV16}, directed motion (DM)~\cite{RUA07,BAN14,MIC10}, anomalous diffusion (AD)~\cite{KNE14} and confined diffusion (CD)~\cite{MON12}. 
The most common analysis method uses mean square displacement (MSD) curves~\cite{MIC10}. Within this approach one fits the theoretical curves for various physical models to the data and then selects the best fit with statistical analysis~\cite{MON12}. However, in many cases the actual trajectories are too short for extracting a meaningful information from the time-averaged MSDs. Moreover, the finite localization precision adds a term to the MSD, which can limit the interpretation of the data~\cite{SAX97,MIC10,KEP15}. Consequently, several alternative methods have been introduced to overcome these problems. For instance, the full distribution of displacements may be fitted to a mixed model in order to extract differences in diffusive behavior between subsets of particle ensembles~\cite{SCH97}. The moment scaling spectrum method can also be used to categorize various modes of motion~\cite{FER01,EWE05}. The distribution of directional changes~\cite{BUR13}, the mean maximum excursion method~\cite{TEJ10} and the fractionally integrated moving average (FIMA) framework~\cite{BUR15} may efficiently replace the MSD estimator for classification purposes. Hidden Markov Models (HMM) has been proposed to check the heterogeneity within single trajectories~\cite{DAS09,SLA15}. They have  proven to be quite useful in the detection of confinement~\cite{SLA18}. Particle filtering may also be used to locate binding sites for the processes with transient confinement~\cite{BER16}.

An alternative approach to an analysis of trajectories rooted in computer science and statistics is also possible. Due to algorithmic advances combined with increased data availability and more powerful computers machine learning (ML) methods may already outperform human experts at some tasks including classification, i.e. the problem of identifying to which category a new observation belongs on the basis of a training dataset containing observations with a known category membership.  Since the detection of the motion falls into the domain of classification, one may try to tackle this problem with machine learning algorithms. This approach is very appealing, because it would enable an automated analysis of many hundreds or even thousands of trajectories with a reduced amount of manual intervention and initial data curation. 

Several attempts to analyze SPT trajectories with ML methods have been already carried out. For instance, Monnier et al~\cite{MON12} used Bayesian approach to MSD-based classification of motion modes. Dosset and coworkers~\cite{DOS16} used a simple back-propagation neural network to discriminate between different types of diffusion. Wagner et al~\cite{WAG17} built a random forest classifier for normal, anomalous, confined and directed diffusion. Although each of these attempts  uses a different machine learning classification algorithm, they all belong to the class of feature-based methods. Each trajectory within this approach is described by a set of human-engineered features and only those features were provided as input to a classifier model. 

In contrast, deep learning methods extract features on their own from raw data, without any effort from human expert. They are gaining on popularity in recent years and were already successfully  applied  to computer vision~\cite{KRI12,SIM14,KAR14}, speech recognition~\cite{DEN13,GRA13} and natural language processing~\cite{COL08,KIM14}. One of the popular methods are convolutional neural networks (CNN)~\cite{LEC98}, which excell in image classification. They have been already applied to single-particle recognition in microscopy experiments~\cite{ZHU17,NEH18}. However, although some attempts to time series analysis with CNNs are already known~\cite{LAN14,QIU14,GAM17}, to the best of our knowledge they have not been applied yet to the problem of classification of motion types from raw trajectories.

Thus, the goal of this paper is to propose a novel approach to SPT trajectory classification based on the CNN deep learning method and to compare its performance with two popular feature-based methods: random forests~\cite{HO95,BRE01} and gradient boosting~\cite{FRI02}. Since all of these methods require large training datasets with trajectories labeled already with a corresponding motion type, we will use synthetic data to train and validate the models. As for the traditional methods, we will follow the approach of Wagner et al~\cite{WAG17} and use their set of features for classification purposes.

The paper is divided as follows. In Sec.~\ref{sec:diff} we introduce basic types of diffusion and briefly discuss the mean square displacement curves as a common tool of trajectory analysis. Classification methods are introduced in Sec.~\ref{sec:class}. In Sec.~\ref{sec:synt} we summarize methods for computer generation of synthetic trajectories. Features used by the traditional classification methods are introduced in Sec.~\ref{sec:feat}. Results of our analysis are presented in Sec.~\ref{sec:resu}, followed by some concluding remarks.

\section{Diffusion modes and their analysis}
\label{sec:diff}

We seek to classify SPT trajectories into four basic motion types: normal diffusion (ND)~\cite{ALV16}, directed motion (DM)~\cite{RUA07,BAN14,MIC10}, anomalous diffusion (AD)~\cite{KNE14} and confined diffusion (CD)~\cite{MON12}. A standard way of identifying them is based on the analysis of the mean square displacement (MSD) of particles~\cite{QIA91}. The MSD is defined as
\begin{equation}
\hat{\rho}(t)\equiv \langle \left(X(t)-X(0)\right)^2\rangle = \frac{1}{M}\sum_{j=1}^M \left( X_j (t) - X_j(0) \right)^2,
\label{eq:msd}
\end{equation}
where $X_j(t)$ is the position of the $j$-th particle after time $t$ and $M$ is the number of particles (i.e. idependent trajectories). MSD is an ensemble average of the square displacement over the probability distribution of $X(t)$. However, due to a limited number of trajectories in many single particle tracking experiments, the ensemble averaged MSD is usually replaced by the time averaged MSD (TAMSD) calculated from a single trajectory. Given a trajectory in form of $N$ consecutive two dimensional positions $X_i =(x_i,y_i)$ ($i=1,\dots,N$) recorded with a constant time interval $\Delta t$, the TAMSD at time lag $n\Delta t$ is defined as
\begin{equation}
\rho(n\Delta t) =\frac{1}{N-n}\sum_{i=1}^{N-n} \left( X_{i+n}-X_i \right)^2.
\label{eq:tamsd}
\end{equation} 
It is worth to mention that for an ergodic process with stationary increments the TAMSD converges to the ensemble averaged MSD in the limit $N\rightarrow\infty$. 

According to Saxton~\cite{SAX97}, for the four basic modes of diffusion we have:
\begin{eqnarray}
\rho_{ND}(n \Delta t) & = & 4Dn\Delta t,\nonumber\\
\rho_{AD}(n \Delta t) & = & 4D(n\Delta t)^\alpha,\label{eq:msds}\\
\rho_{DM}(n \Delta t) & = & 4Dn\Delta t + (vn\Delta t)^2,\nonumber\\
\rho_{CD}(n \Delta t) & \simeq & r_c^2 \left[1-A_1\exp\left(\frac{-4A_2Dn\Delta t}{r_c^2} \right)\right].\nonumber
\end{eqnarray} 
Here, $\alpha<1$ is the anomalous exponent, $v$ is the velocity in the directed motion, the constants $A_1$ and $A_2$ characterize the shape of the confinement and $r_c$ is the confinement radius.

For pure trajectories with no localization errors one could actually determine their diffusion modes  simply based on the shapes of MSD curves and their mathematical models given by Eq.~(\ref{eq:msds}). However, in case of real trajectories there is usually a lot of noise in the data, which makes the fitting of a mathematical model a challenging task, even in the simplest case of the normal diffusion~\cite{MIC10}. Moreover, according to Eq.~(\ref{eq:tamsd}), only the MSD values corresponding to small time lags are well averaged. The larger the lag the smaller is the number of displacements contributing to the averages, resulting in fluctuations increasing with the lag. This constitutes a problem in particular in case of short trajectories, for which the fit to mathematical models has to be limited to just a few first time lags. This is the reason why we are interested in classification methods that go beyond fitting of mathematical models to the MSD curves.

\section{Classification methods}
\label{sec:class}

Traditional machine learning is a set of methods of statistical learning where each instance in a dataset is described by a set of human-engineered features or attributes~\cite{MIT97}. In contrast, deep learning methods extract features from raw data  without any effort from human expert~\cite{HAT18}. The representation of data is constructed  automatically and there is no need for complex data preprocessing as in the case of the machine learning.

The deep learning approach constitutes nowadays the state-of-the-art technology for automatic data classification and overshadows a little bit the classical machine learning algorithms. However, in some specific situations the latter ones are still better to use. The reasons are at least threefold: they work better on small data, are financially and computationally cheaper and usually are easier to interpret. Thus the ultimate goal of this paper is to compare the performance of machine and deep learning algorithms applied to the recognition of the diffusion type in single particle tracking data. We will examine two classical algorithms, i.e. random forests~\cite{HO95,BRE01} and gradient boosting~\cite{FRI02}, together with convolutional neural networks (CNN)~\cite{LEC98}.

\subsection{Feature-based methods}

Both random forests and gradient boosting algorithms belong to the class of ensemble learning, i.e. methods that generate many classifiers and aggregate their results. In both cases, decision trees~\cite{SON15} are used as the basic classifier.

Decision trees are used very often for classification purposes, because they are easy to understand and interpret. And they usually do not require a data preprocessing. However, they are unstable in the sense that a small variation in the data may lead to a completely different tree~\cite{GAR13}. And they have the tendency to overfit, i.e. they correspond to closely or exactly to a particular set of data, and may therefore fail to fit additional data or predict future observations reliably~\cite{BRA13}.  Although methods like prunning are known to avoid overfitting, it is the main reason why the decision trees are used as building blocks of ensemble classifiers rather than standalone ones.

\subsubsection{Random forests}

In a random forest, several decision trees are constructed from the same training data. For a given input, the predictions of individual trees are aggregated and then their mode is outputted as the class of the input data.
A modern version of the algorithm combines the bagging idea proposed by Breiman~\cite{BRE01} with
the random subspace method invented by Ho~\cite{HO95,HO98}. Bagging repeatedly selects a random sample with replacement of the training set and fits trees to these samples. In order to avoid correlations between the trees, for each one a random subset of features is selected. Typically, in a classification problem with $N$ features, $\sqrt{N}$ of them are used to build one tree.

\subsubsection{Gradient boosting}

\begin{figure}
\includegraphics[scale=1]{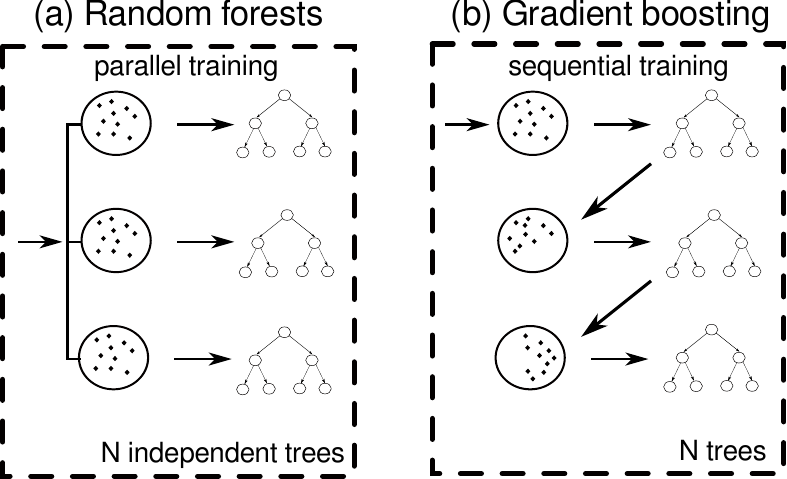} %boosting.pdf
\caption{Comparison between (a) random forest and (b) gradient boosting methods. In the random forest, $N$ independent learners (trees) are built in parallel from random subsets of the input data set. In gradient boosting, the next tree is  constructed from the pseudo-residuals of the ensemble and added to it.\label{boosting}}
\end{figure}

In contrast to random forests, the trees in gradient boosting are not independent. Instead, the single classifiers are built sequentially by learning from mistakes committed by the ensemble~\cite{SCH98,FRI02} (see Fig.~\ref{boosting} for a schematic comparison of the two methods).

\subsection{Deep learning methods}
\label{sec:dl}

Deep learning (DL) methods operate on raw data. They do not require any feature selection and extraction carried out by a human expert. Instead, they use a cascade of multiple layers of nonlinear processing units for feature identification, extraction and transformation in order to learn multiple levels of data representations~\cite{DEN14}. 

\begin{figure}
\includegraphics[scale=1]{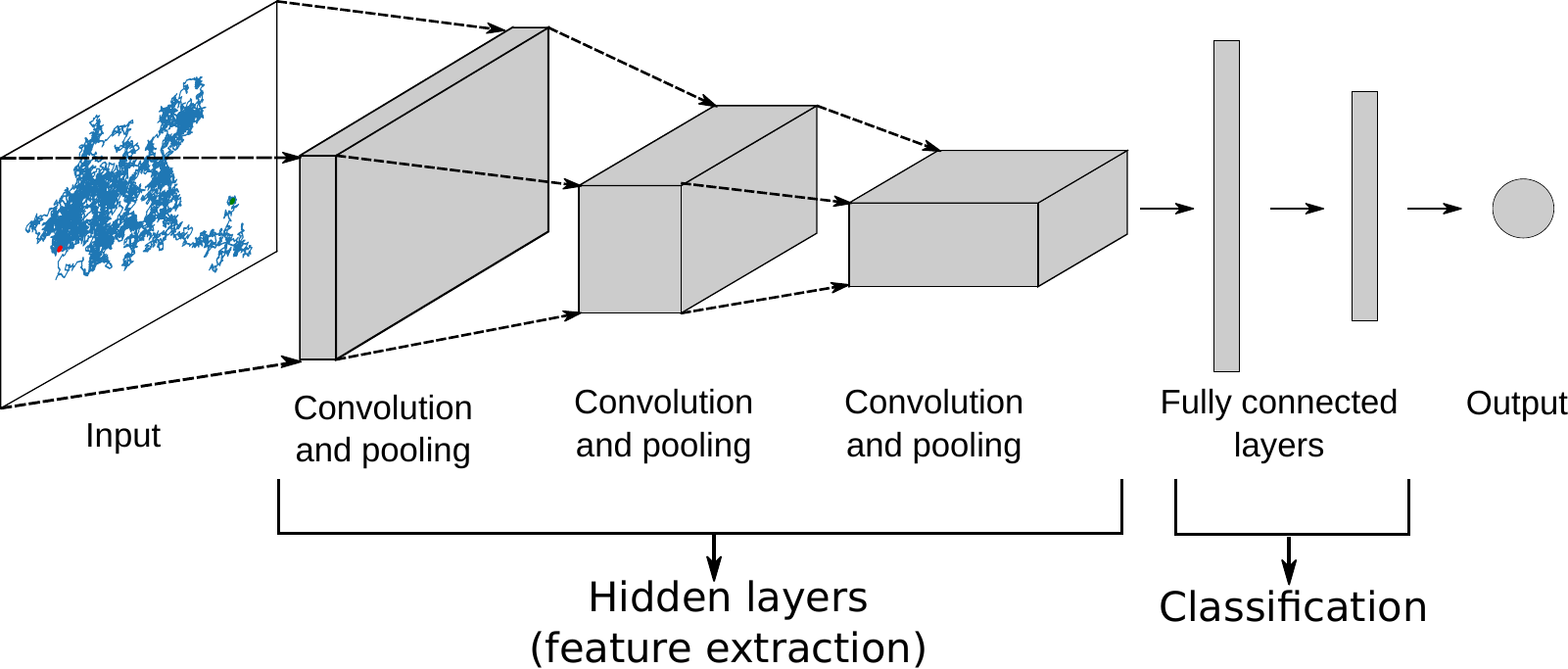} %cnn.pdf
\caption{A schematic architecture of a CNN. The network consists of two components: hidden layers responsible for feature extraction from input data and fully connected layers carrying out the classification.\label{cnn}}
\end{figure}

In this paper, we are going to use convolutional neural networks for trajectory classification. They have been already successfully applied to many tasks including a  time series analysis~\cite{YAN15}. A schematic architecture of a CNN is shown in Fig.~\ref{cnn}. Such a network has usually two components. The one consisting of hidden layers is responsible for extraction of features from the raw input data. The layers will perform a series of convolutions and pooling operations during which attributes of data are detected.  Each convolution uses a different filter which is sliding over the input and producing its own feature map in form of a 3D array. All the maps are then combined together as the final output of the component. The role of pooling is to reduce the dimensionality of feature maps in order to decrease the number of parameters and computations in the network.
The classification part contains few fully connected layers like in a regular neural network~\cite{GAR98}. Flattening of data is usually required at the interface between the components, because the fully connected layers can process only 1D vectors.

\section{Synthetic data}
\label{sec:synt}

All three methods described in the previous section belong to the class of supervised learning, i.e. they infer a model from a set of training examples~\cite{RAS15}. Each sample is a pair consisting of an input object (a trajectory) and a desired output value (a diffusion mode). The model is a function that maps an input to an output and can be used for classification of new input data.

Since thousands of labeled trajectories are needed to train the classifiers, especially in the deep learning case, we will use computer-generated synthetic 2D trajectories as our training set. Simulation methods for every type of diffusion will be briefly discussed below.

\subsection{Normal diffusion}

According to Michalet~\cite{MIC10}, the probability distribution of the displacement's norm in case of the normal diffusion is given by
\begin{equation}
F_d(u) = \frac{2u}{4D\Delta t} \exp \left( \frac{-u^2}{4D\Delta t} \right),~u\geq 0
\label{eq:rayleigh}
\end{equation}
where $\Delta t$ is the time interval during which the displacement is recorded. Mathematically, Eq.~(\ref{eq:rayleigh}) is  a Rayleigh distribution~\cite{PAP02}. To simulate a trajectory, we randomly choose a start position of a particle and a random direction of displacement $\alpha$ and then pick up a random steplength $d$ from Eq.~(\ref{eq:rayleigh}). Then we calculate the new position of the particle and take it as the starting point for the next step. The procedure is repeated a given number of times.

\subsection{Directed motion}

Once we have a procedure generating a normal diffusion trajectory, simulation of the directed motion is straightforward. For a given velocity $\vec{v}$, in each step we simply calculate a correction to the position due to the active motion,
\begin{eqnarray}
dx_i & = & v\Delta t\cos\beta ,\label{eq:positions2}\\
dy_i & = & v\Delta t\sin\beta ,\nonumber
\end{eqnarray}
and add it to the new coordinates:
\begin{eqnarray}
x_{new} & = & x_{old} + d\cos\alpha +dx_i,\label{eq:positions3}\\
y_{new} & = & y_{old} + d\sin\alpha +dy_i.\nonumber
\end{eqnarray}
The angle $\beta$ in Eq.~(\ref{eq:positions2}) is the direction of the velocity.

Following Wagner et al~\cite{WAG17}, we may want to introduce a measure of how a trajectory is influenced by the active motion,
\begin{equation}
R=\frac{v^2 T}{4D},\label{eq:r}
\end{equation}
with $T$ being the time duration. This measure can be helpful in generating similar trajectories with different values of $v$ and $D$.

\subsection{Confined diffusion}

A small modification of the normal diffusion procedure is needed to simulate a confined diffusion~\cite{WAG17}. We assume that a particle starts from the center of a 2D circular reflective boundary. We divide every step of the simulation in 100 substeps with $\Delta t^\prime=\Delta t/100$. In every substep we carry out a normal diffusion step. The position of the particle after the substeps will be updated only if the distance from the center to new coordinates is smaller than the radius of the reflective boundary.

Wagner et al~\cite{WAG17} have introduced the boundedness parameter $B$, defined as the area of the smallest ellipse enclosing a normal diffusion trajectory (with no confinement) divided by the area of the confinement,
\begin{equation}
B=\frac{A_{ellipse}}{\pi r_c^2}\simeq\frac{DN\Delta t}{ r_c^2}.\label{eq:b}
\end{equation}
As in the case of the directed motion, this parameter will help to evaluate trajectories independently of the actual values of $D$ and $r_c$. 

\subsection{Anomalous diffusion}

Anomalous diffusion was simulated with the fractional Brownian motion (FBM)~\cite{MAN68}. 
FBM is a continuous-time  Gaussian process $B_H (t)$ on $[0,T]$, that starts at zero, has expectation zero for all $t\in [0,T]$ and its covariance function is given by
\begin{equation}
E\left[ B_H(t)B_H(s) \right]=\frac{1}{2}\left( |t|^{2H} + |s|^{2H}-|t-s|^{2H}\right),
\end{equation}
where the Hurst index $H$ is a real number in $(0,1]$. There is a simple realtion between $H$ and the anomalous diffusion exponent $\alpha$ introduced in Eq.~(\ref{eq:msds}), namely
\begin{equation}
2H = \alpha.
\end{equation}

The value of the Hurst index determines the type of motion generated by the process. $H=1/2$ (i.e. $\alpha=1$) corresponds to a normal diffusion. For $H>1/2$ ($\alpha>1$), the increments of FBM are positively correlated, resulting in a super-diffusion. Finally, negative correlations between FBM increments occur for $H<1/2$ ($\alpha<1$), leading to a sub-diffusion. We focus on the latter case in this work.

We used a dedicated Python package called \texttt{fbm} to simulate the fractional Brownian motion~\cite{FLY17}. Davies-Harte algorithm~\cite{DAV87} was utilized to generate independent trajectories of the process.

\subsection{Adding noise}

Real trajectories can be altered by various measurement noises such as localization errors, electronic noise, drift or vibrations of the sample or postprocession errors~\cite{LAN18}. To account for these issues, we added normal Gaussian noise with zero mean and the standard deviation $\sigma$ to each simulated position.

Let us first introduce two different signal to noise ratios (SNR): one for ND, AD and CD,
\begin{equation}
Q_1 = \frac{\sqrt{D\Delta t}}{\sigma}\label{eq:snr1},
\end{equation}
an another one for DM,
\begin{equation}
Q_2 = \frac{\sqrt{D\Delta t +v^2\Delta t^2}}{\sigma}\label{eq:snr2}.
\end{equation}
Instead of setting $\sigma$ directly in our simulations, we will prefer to set a random level of $SNR$ first and then 
to calculate the standard deviation for given $D$ and $\Delta t$ from one of the above equations.

\subsection{Simulation details}
\label{sec:sim_details}

Our training data consists of 20000 synthetic trajectories, i.e. 5000 for each diffusion type. Following Wagner et al~\cite{WAG17} we used fixed values for two of the parameters: $\Delta t =1/30~s$ and $D=9.02~\mu m^2/s$. As for the timelag, it is a typical value in experimental setups. The value of the diffusion coefficient $D$ corresponds to a freely diffusing nanoparticle with a diameter $50~nm$ in water at $22^\circ C$. Other parameters were chosen randomly. Their values are summarized in Table~\ref{tab:parameters}. We used our own codes written in Python to simulate the training set. The codes are available upon request.
\begin{table}
\begin{tabular}{c|l|c}
\hline\hline
Parameter & Meaning & Range of values\\
\hline
$\Delta t$ & timelag between steps & $1/30~[s]$ \\
$D$ & diffusion coefficient & $9.02~[\mu m^2/s]$\\
$N$ & length of a trajectory & $30-600$ \\
$B$ & boundedness& $1-6$\\
$R$ & active motion to diffusion ratio & $1-17$ \\
$\alpha$ & anomalous exponent & $0.3-0.7$ \\
$Q$ & signal to noise ratio & $1-9$\\
\hline\hline
\end{tabular}
\caption{Parameters of the simulation and their values. All parameters except $\Delta t$ and $D$ were randomly chosen from given ranges.\label{tab:parameters}}
\end{table}

\section{Feature extraction}
\label{sec:feat}

We will follow the approach of Wagner et al~\cite{WAG17} and use their nine features together with the diffusion coefficient fitted from the data as the tenth one. In this section we will give a short description of the features used for training of our classifiers.

\subsection{Diffusion coefficient}

We will use the diffusion coefficient of the model given by the first of Eqs.~(\ref{eq:msds}) fitted to the mean square displacement curve estimated by Eq.~(\ref{eq:tamsd}).

\subsection{Anomalous exponent}

Anomalous exponent $\alpha$ is the exponent in the second model defined in  Eqs.~(\ref{eq:msds}). Again, it will  be fitted to the MSD curve obtained from Eq.~(\ref{eq:tamsd}).

\subsection{Asymmetry}

The asymmetry of a trajectory can be used to detect directed motion. Following Saxton~\cite{SAX93} we will derive it from the gyration tensor, which describes the second moments of positions of a particle. For a 2D random walk of $N$ steps it is given by
\begin{equation}
\mathbf{T} =\left(
\begin{array}{cc}
\frac{1}{N}\sum_{j=1}^N (x_j -\langle x \rangle)^2 & \frac{1}{N}\sum_{j=1}^N (x_j -\langle x \rangle)(y_j -\langle y \rangle) \\ 
\frac{1}{N}\sum_{j=1}^N (x_j -\langle x \rangle)(y_j -\langle y \rangle) & \frac{1}{N}\sum_{j=1}^N (y_j -\langle y \rangle)^2
\end{array} 
\right), \label{eq:tensor}
\end{equation} 
where $\langle x \rangle=(1/N)\sum_{j=1}^N x_j$ is the average of $x$ coordinates over all steps in the random walk. We will define the asymmetry as~\cite{HEL07}
\begin{equation}
A=-\log \left(1 - \frac{(\lambda_1-\lambda_2)^2}{2(\lambda_1+\lambda_2)} \right),\label{eq:asymmetry}
\end{equation}
where $\lambda_1$ and $\lambda_2$ are the principle radii of gyration, i.e. the eigenvalues of the tensor $\mathbf{T}$.

\subsection{Efficiency}

Efficiency relates the net squared displacement of a particle to the sum of squared step lengths,
\begin{equation}
E = \frac{|X_{N-1}-X_0|^2}{(N-1)\sum_{i=1}^{N-1}|X_i -X_{i-1}|^2}. \label{eq:efficiency}
\end{equation} 
It is a measure for linearity of a trajectory and like asymmetry, it may help to detect directed motion.

\subsection{Fractal dimension}

The fractal dimension is a measure of the space-filling capacity of a pattern. According to Katz and George~\cite{KAT85}, the fractal dimension of a trajectory can be calculated as
\begin{equation}
D_f =\frac{\log N}{\log(NdL^{-1})},\label{eq:fd}
\end{equation}
where $L$ is the total length of the path, $N$ is the number of steps and $d$ is the largest distance between any two positions.

The measure takes values around 1 for straight trajectories (direct motion), around 2 for random ones (normal diffusion) and around 3 for constrained trajectories (confined or anomalous diffusion)~\cite{KAT85}.

\subsection{Gaussianity}

Trajectory's gaussianity was introduced by Ernst et al~\cite{ERN14} to check the Gaussian statistics on increments,
\begin{equation}
g(n) = \frac{\langle r_n^4 \rangle}{2\langle r_n^2 \rangle^2},\label{eq:gaussianity}
\end{equation}
where the trajectory's quatric moment is given by
\begin{equation}
\langle r_n^4 \rangle = \frac{1}{N-n}\sum_{i=1}^{N-n} |X_{i+n}-X_i|^4.
\end{equation}
For normal diffusion we should get gaussianity equal to 0. Since we used FBM, which has Gaussian increments, to simulate anomalous diffusion, we expect to get the same result for AD. The other types of motion should show deviations from 0.

\subsection{Kurtosis}

Kurtosis measures the asymmetry and peakiness of the distribution of points within a trajectory~\cite{HEL07}. For its calculation the position vectors $X_i$ are projected onto the dominant eigenvector $\vec{r}$ of the gyration tensor~(\ref{eq:tensor}) yielding scalars
\begin{equation}
x_i^p = X_i\cdot\vec{r}.
\end{equation}
Kurtosis is defined as the fourth moment of the set of $x_i^p$,
\begin{equation}
K=\frac{1}{N}\sum_{i=1}^N \frac{(x_i^p -\bar{x}^p)^4}{\sigma^4_{x^p}},\label{eq:kurtosis}
\end{equation}
with $\bar{x}^p$ being the mean projected position and $\sigma_{x^p}$ - the standard deviation of $x^p$.

\subsection{MSD ratio}

The mean square displacement ratio characterizes the shape of the MSD curve. We will define it as follows:
\begin{equation}
\kappa(n_1,n_2) = \frac{\langle r_{n_1}^2 \rangle}{\langle r_{n_2}^2 \rangle} - \frac{n_1}{n_2},\label{eq:msdr}
\end{equation}
where $n_1<n_2$. Taking Eq.~(\ref{eq:msds}) into account we see that $\kappa=0$ for normal diffusion, it is negative for direct motion and positive for other types of diffusion. In our analysis we simply took $n_2 = n_1 + \Delta t$ and calculated an averaged ratio for every trajectory. 

\subsection{Straightness}

Straightness is a measure of the average direction change between subsequent steps. Similar to efficiency it relates the net displacement to the sum of step lengths:
\begin{equation}
S = \frac{|X_{N-1}-X_0|}{\sum_{i=1}^{N-1}|X_i -X_{i-1}|}. \label{eq:straightness}
\end{equation}

\subsection{Trappedness}

Trappedness is the probability that a diffusing particle with the diffusion coefficient $D$ and traced for a time interval $t$ is trapped in a bounded region with radius $r_0$. According to Saxton~\cite{SAX93} it can be estimated by
\begin{equation}
P(D,t,r_0) = 1-\exp\left( 0.2045 -0.25117\left( \frac{Dt}{r_0^2} \right)  \right).\label{eq:trappedness}
\end{equation}
Since the radius $r_0$ is usually not known, we will approximate it by the half of the maximum distance between any two positions along a given trajectory. For $D$, we will take its short-time estimate, fitted to the first two points of the MSD curve.

\section{Results}
\label{sec:resu}

We decided to use existing machine learning libraries within this project. Random forest and gradient boosting implementations available in \texttt{scikit-learn}~\cite{PED11}, the most popular ML learning library in Python, were used to build the feature-based classifiers. And we used \texttt{mcfly}~\cite{KUP17}, a deep learning library for time series processing, to find and train a deep classifier working with raw diffusion data. All codes were written in Python and are available on request. The computations were carried out on a cluster of 24 CPUs (2.6 GHz each) with the total memory of 50 GB. If not stated otherwise, the synthetic trajectories were randomly split into two subsets: a training set containing 70\% of them and a test set.

\subsection{Featured-based classification}

The random forest classifier implemented in \texttt{scikit-learn} follows the original paper by Breiman~\cite{BRE01}. The gradient boosting algorithm available in this library is described in Refs.~\cite{FRI99,FRI02}. The parameters of the models were optimized with a randomized search method (the \texttt{RandomizedSearchCV} function in \texttt{scikit-learn}). They are summarized in Table~\ref{tab:best_model_parameters}.

\begin{table}
\begin{tabular}{p{7cm}|c|c}
\hline \hline
 & Random Forest & Gradient boosting \\ 
\hline 
number of trees & 500 & 500 \\ 
\hline 
maximum depth of a single tree & 20 & 10 \\ 
\hline 
minimum number of samples required to split an internal node & 2 & 5 \\ 
\hline 
minimum number of samples required to be at a leaf node & 1 & 4 \\ 
\hline\hline 
\end{tabular} 
\caption{Optimal parameters for the random forest and gradient boosting models trained on our data. A randomized search method was used to determine those values. \label{tab:best_model_parameters}}
\end{table}

\subsubsection{Accuracy}

One of the basic metrics used to asses the performance of classification models is accuracy, defined as the number of correct predictions divided by the total number of predictions.

In Table~\ref{tab:accuracy}, accuracies for both feature-based classifiers are shown. The numbers in the first row correspond to the accuracy achieved after a single random split into training and test sets. For the second row we used the 10-fold cross validation method. The idea behind this technique is to randomly split the data set into 10 folds without replacement and use 9 of them for training and one for testing of the model. The procedure is repeated 10 times, so we obtain 10 models and accuracy estimates. An average of those estimates gives the overall accuracy.
\begin{table}
\begin{tabular}{c|c|c}
\hline \hline
 & Random forest & Gradient boosting \\ 
\hline 
Single split of data & 96.43\% & 96.97\% \\ 
\hline
10-fold cross validation & 96.23\% & 96.47\% \\ 
\hline \hline
\end{tabular} 
\caption{Accuracies of the feature-based classifiers. \label{tab:accuracy}}
\end{table}

Since in the gradient boosting an ensemble of the decision trees is built with the purpose to reduce the total error, we would expect that the algorithm performs much better than the random forest. From Table~\ref{tab:accuracy} it follows, that its accuracy is indeed higher, but the differences are actually negligible. Both classifiers perform excellent with an average accuracy of more than 96\%. 

\begin{figure}
\includegraphics[scale=0.5]{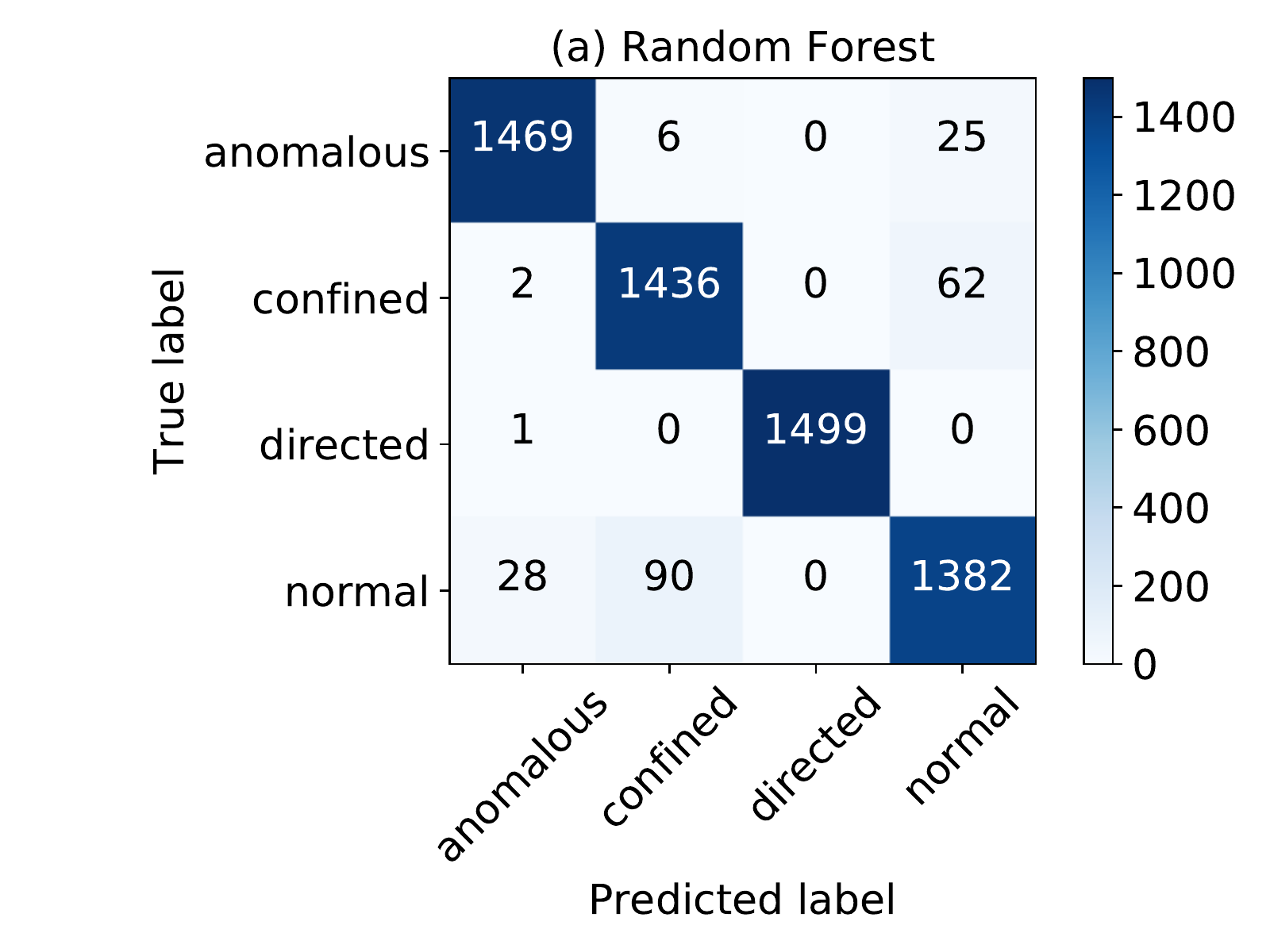}\ %confusion_matrix_rf.pdf
\includegraphics[scale=0.5]{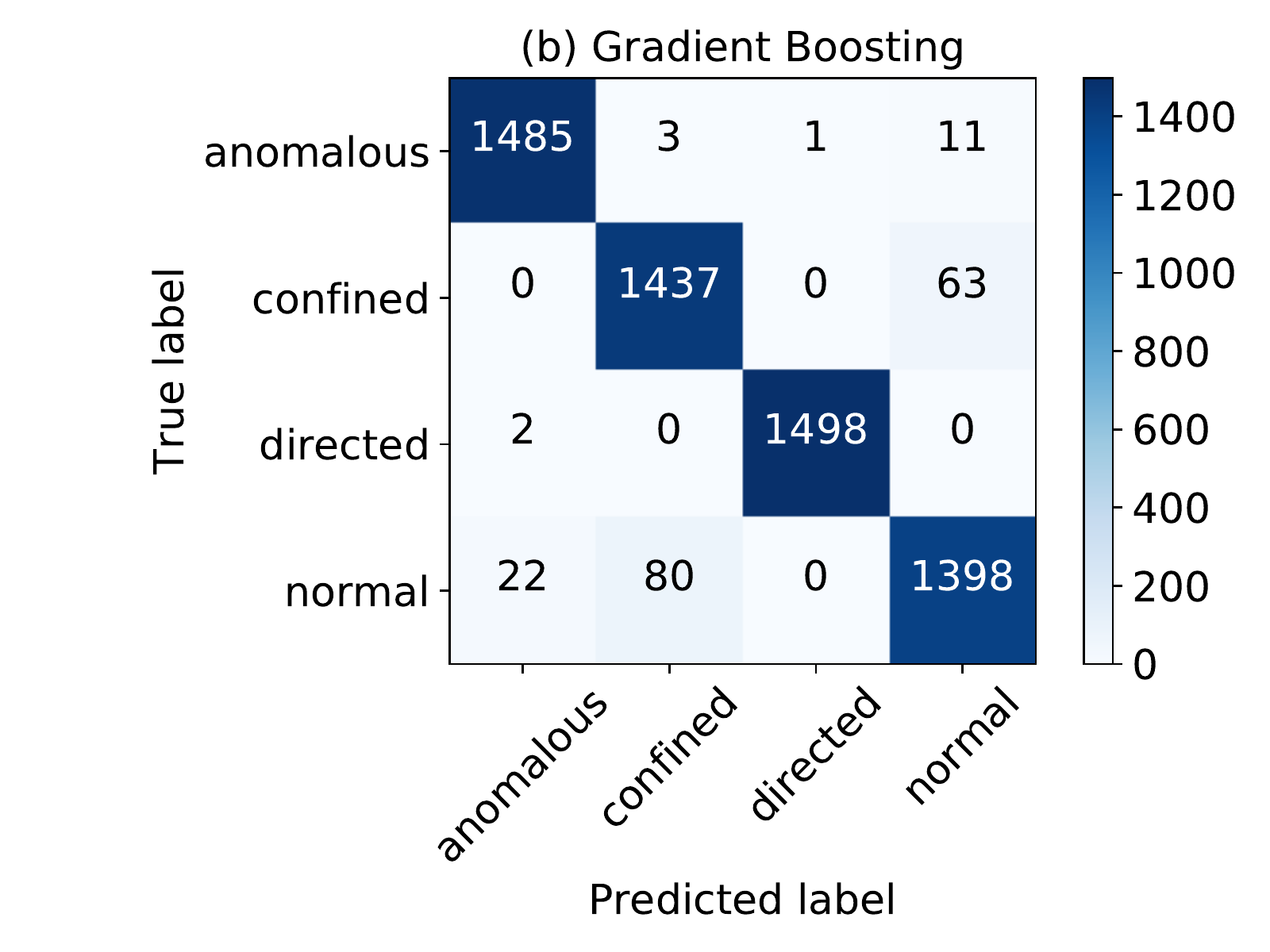}  %confusion_matrix_gb.pdf
\caption{Confusion matrices for (a) the random forest  and (b) the gradient boosting classifiers.\label{fig:cm}}
\end{figure}

In Fig.~\ref{fig:cm}, confusion matrices of the classifiers are presented. 
In both cases the classifiers made a total of 6000 predictions (sum of all matrix elements), including 1500 for each type of diffusion (sum of all elements in a matrix row). As far as the random forest model is concerned, the best performance was observed for the directed diffusion - among the 1500 directed trajectories only one was wrongly classified as an anomalous one. The performance decays slightly for the anomalous and confined modes and is significantly worse for the normal diffusion. The gradient boosting model reveals similar characteristics with slightly different absolute numbers.

The data collected in the confusion matrices may be used to generate a more detailed description of the performance of the models under investigation. The results are briefly summarized in Table~\ref{tab:report_fb}.
\begin{table}
\begin{tabular}{c||c|c|c||c|c|c}
\hline \hline
 & \multicolumn{3}{c||}{Random forest} & \multicolumn{3}{|c}{Gradient boosting} \\ 
\hline 
 & precision & recall &  support & precision & recall &  support \\ 
\hline 
anomalous & 0.98 & 0.98 &  1500 & 0.98 &  0.99  &   1500 \\ 
\hline 
confined & 0.94 & 0.96 &  1500 & 0.94  &  0.96 &    1500\\ 
\hline 
directed & 1.000 & 1.00 &  1500 & 1.00  &  1.00 &    1500\\ 
\hline 
normal & 0.94 & 0.92 &  1500 & 0.95  &  0.93 &    1500\\ 
\hline 
average/total & 0.96 & 0.96 &  6000 & 0.97  &  0.97  &   6000\\ 
\hline 
accuracy & \multicolumn{3}{c||}{96.43\%} & \multicolumn{3}{c}{96.97\%} \\ 
\hline \hline
\end{tabular} 
\caption{A brief summary of the performance of feature-based classifiers. All results are rounded to two decimal digits.\label{tab:report_fb}}
\end{table}
Here, we adopted two quantities commonly  used in classification problems: precision and recall~\cite{PER55}. Precision is the fraction of correct predictions among all predictions. It tells us how often a classifier is correct if it predicts a given class. Recall is the fraction of correct predictions of a given class over the total number of members of this class. Despite small differences in the numbers, each of our models is characterized by both very high precision and recall. Thus, they not only return much more relevant results than the irrelevant ones (high precision), but also yield most of the relevant results (high recall).

\subsubsection{Feature importances}

A nice detail of the ensemble classification methods is that they usually allow one to easily compute the relative importances of features for a given problem. Variables with high importance scores are the drivers of the outcome and their values have a significant impact on the correctness of a  prediction. Features with low importance might be usually omitted from a model, making it faster to fit and predict. 

\begin{figure}
\includegraphics[scale=0.5]{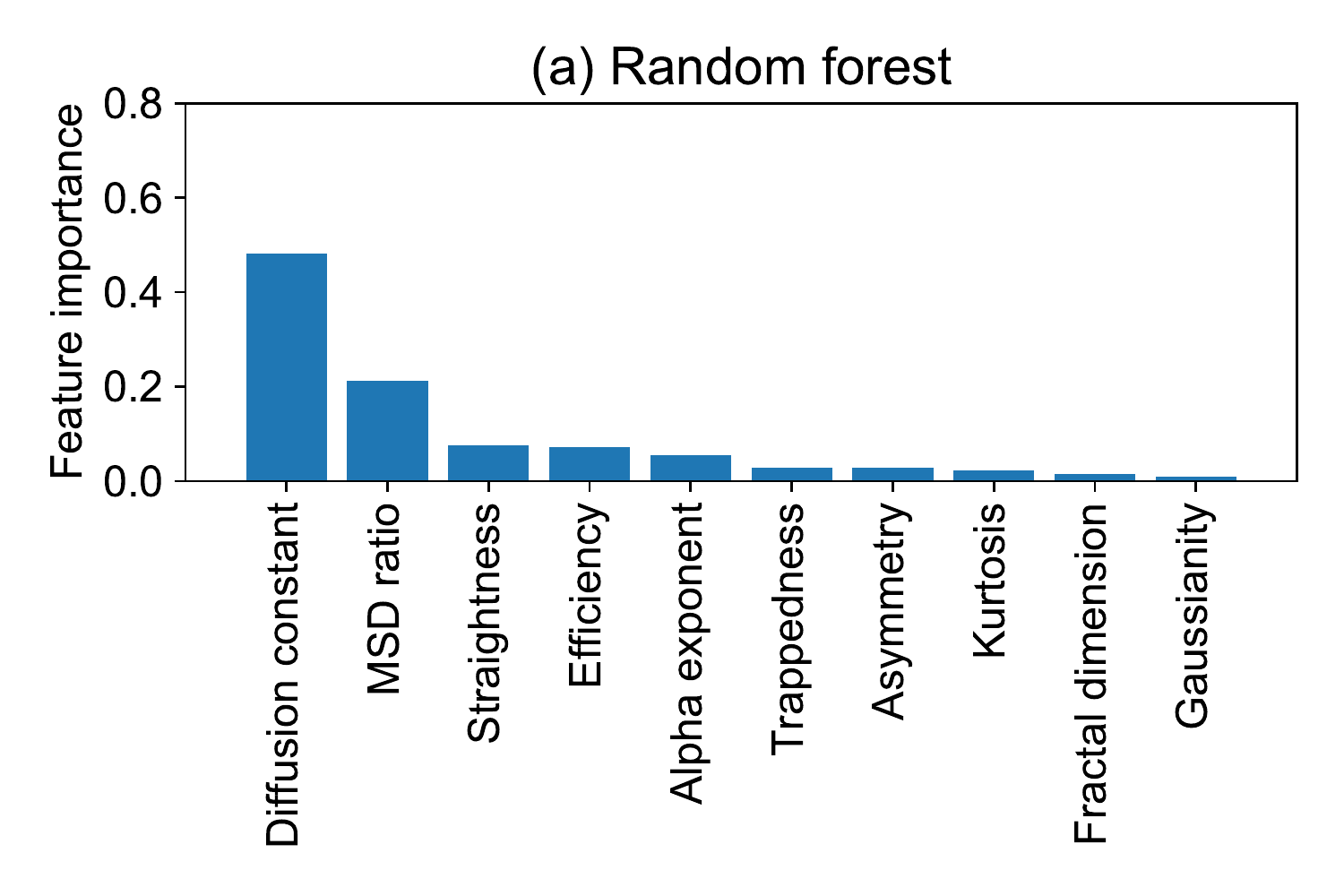}\ % feature_importances_rf.pdf
\includegraphics[scale=0.5]{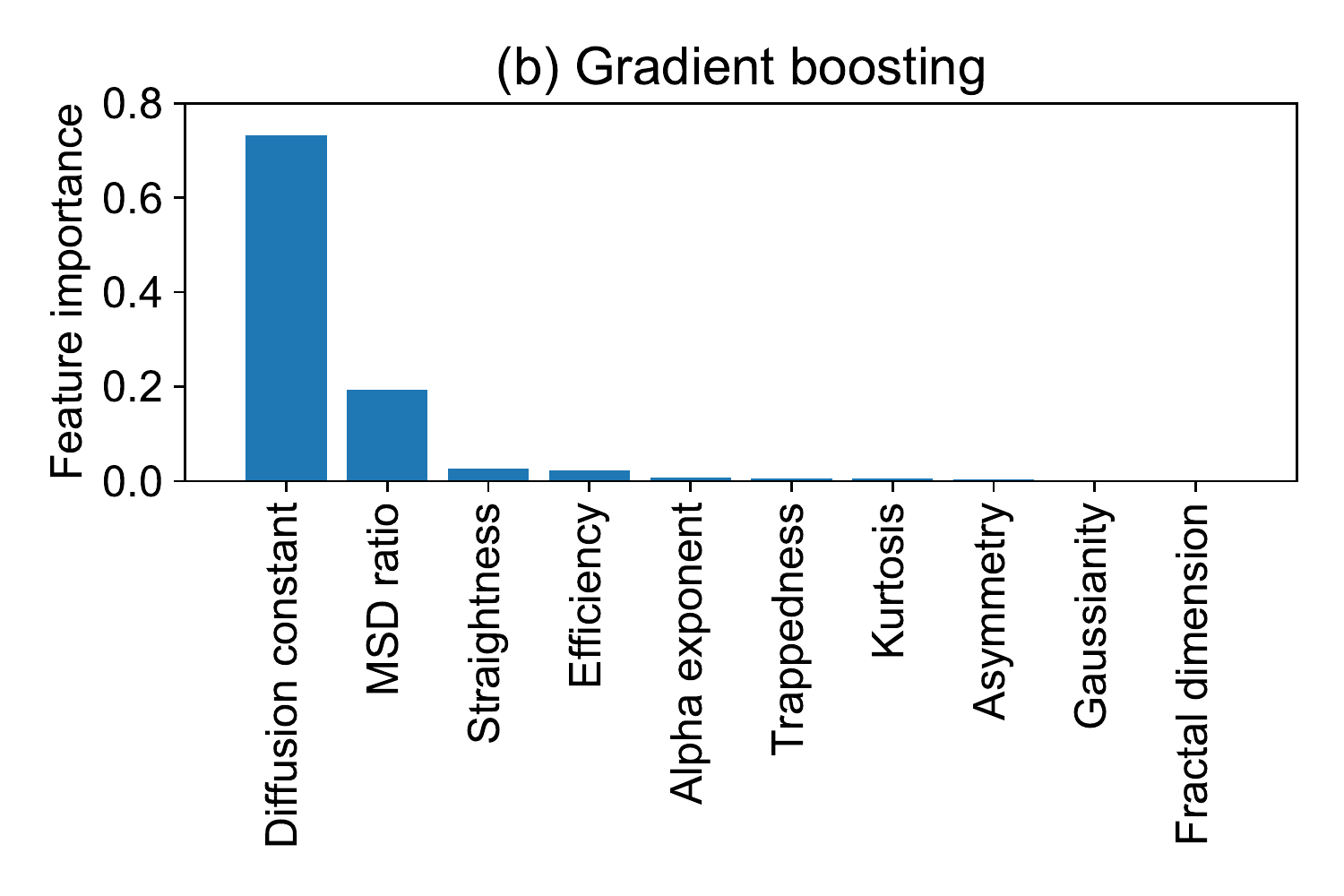}  % feature_importances_gb.pdf
\caption{Feature importances in (a) the random forest and (b) the gradient boosting models.\label{fig:importances}}
\end{figure}

The \texttt{scikit-learn} implementations of the random forest and gradient boosting classifiers calculate the importances on the fly during the training process and provide an interface to access them. Results are shown in Fig.~\ref{fig:importances}.
The linear diffusion coefficient $D$ seems to be the most important feature in both cases, followed by the MSD ratio, straightness, efficiency and the anomalous exponent $\alpha$. There are some differences between the methods as well. For instance, the dominance of the diffusion coefficient over all other features is less pronounced in the random forest. Instead, we observe  non-vanishing importances of the remaining features, i.e. trappedness, asymmetry, kurtosis, fractal dimension and gaussianity. In contrast, the importances of those features are negligible in the gradient boosting case and the difference between the first and the second rank  is much higher.

\begin{figure}
\includegraphics[scale=0.5]{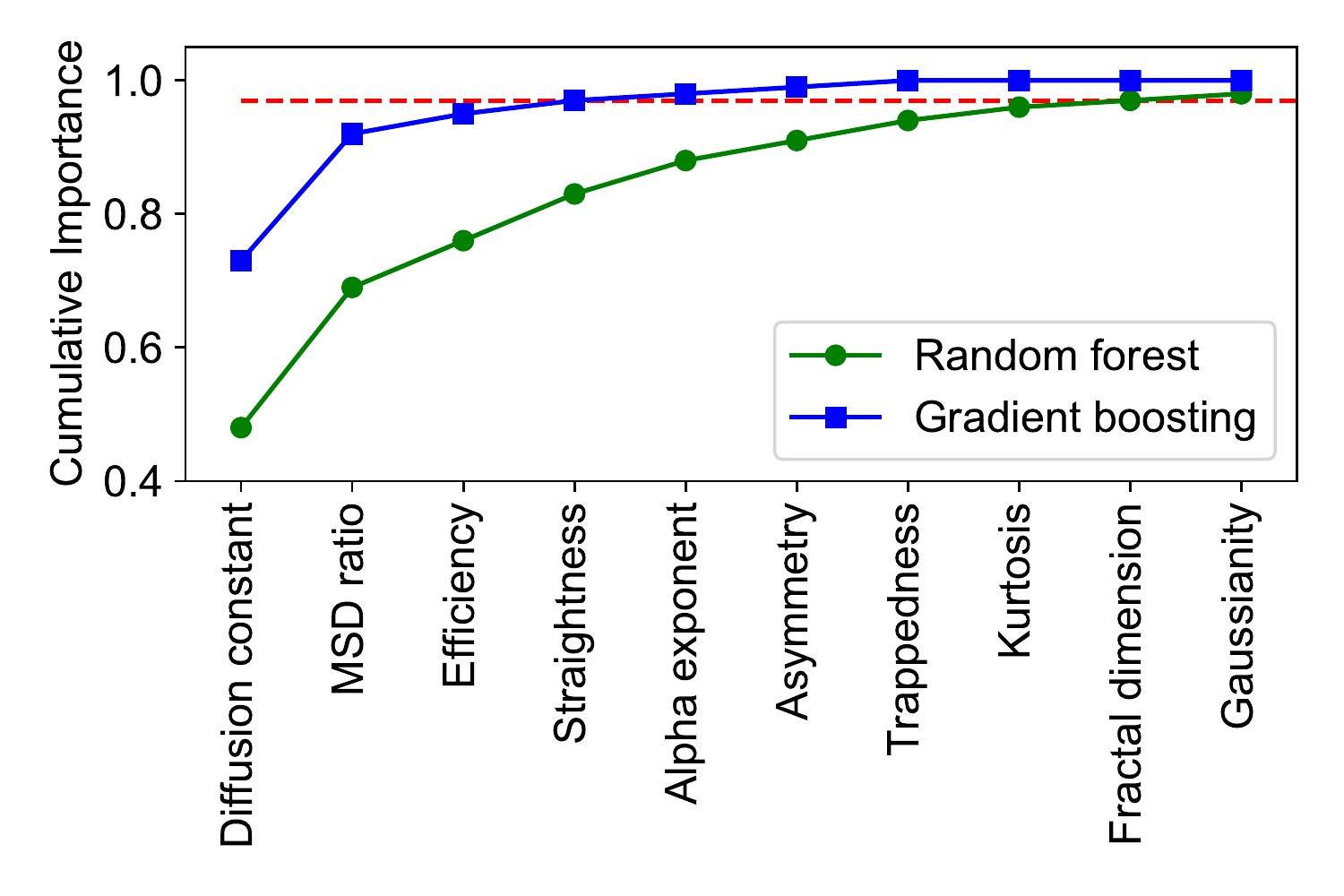} %cumulative_importances.pdf
\caption{Cumulative importance of features for both models. The dashed line is the 97\% level of importance and indicates a threshold for feature selection.\label{fig:cum_imp}}
\end{figure}

To illustrate the differences between the models, in Fig.~\ref{fig:cum_imp} we show the cumulative importances of features. The dashed horizontal line in this plot is the 97\% level of importance and could indicate a threshold for feature selection, i.e. once the level is reached, we can omit the remaining features without affecting the model very much. In order to find a value of the threshold, one should check how his model generalizes to unseen data after removing attributes for different thresholds and then chose the one not negatively affecting the accuracy of the model.

\begin{figure}
\includegraphics[scale=0.5]{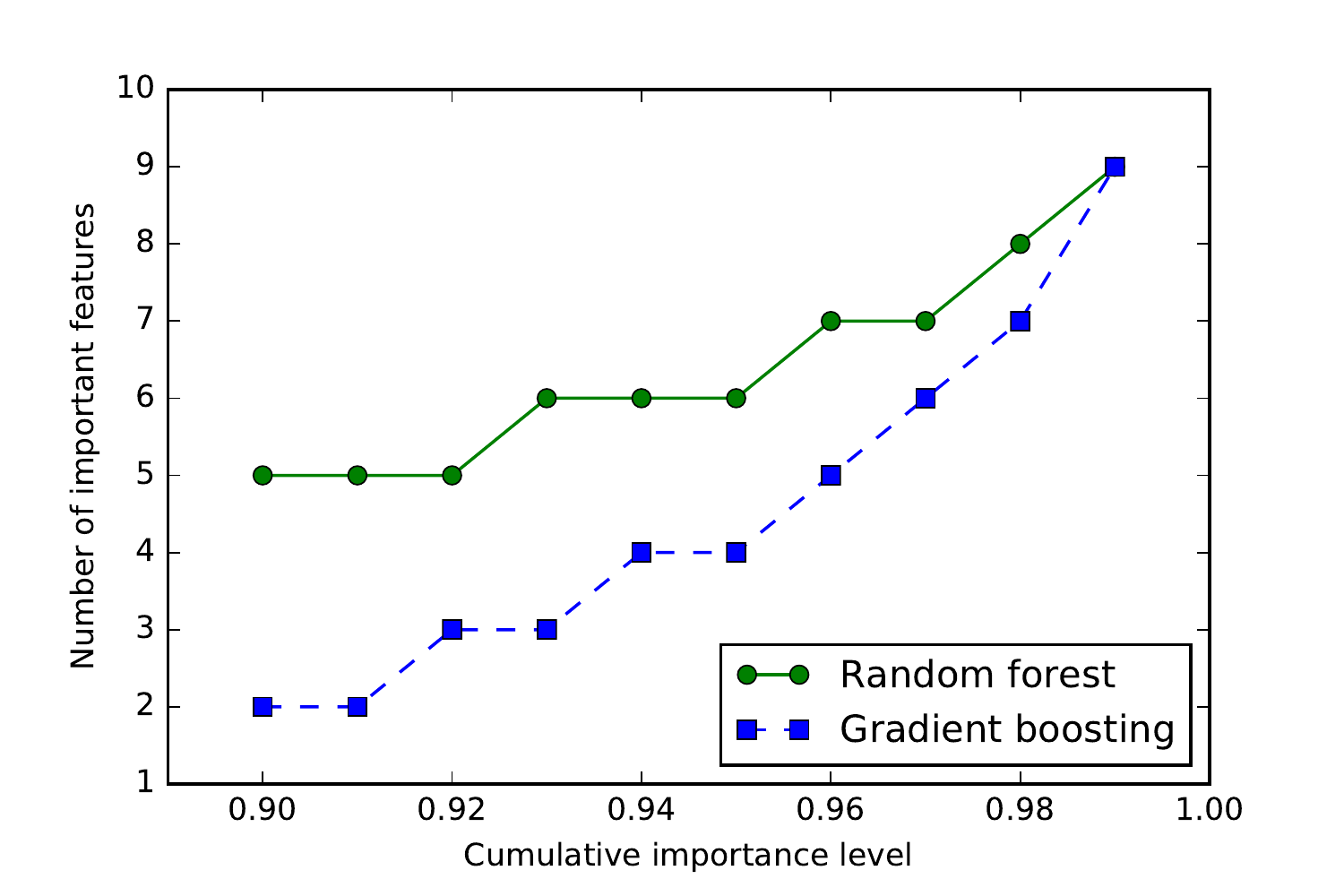}\ %cumulative_importance_threshold.pdf
\includegraphics[scale=0.5]{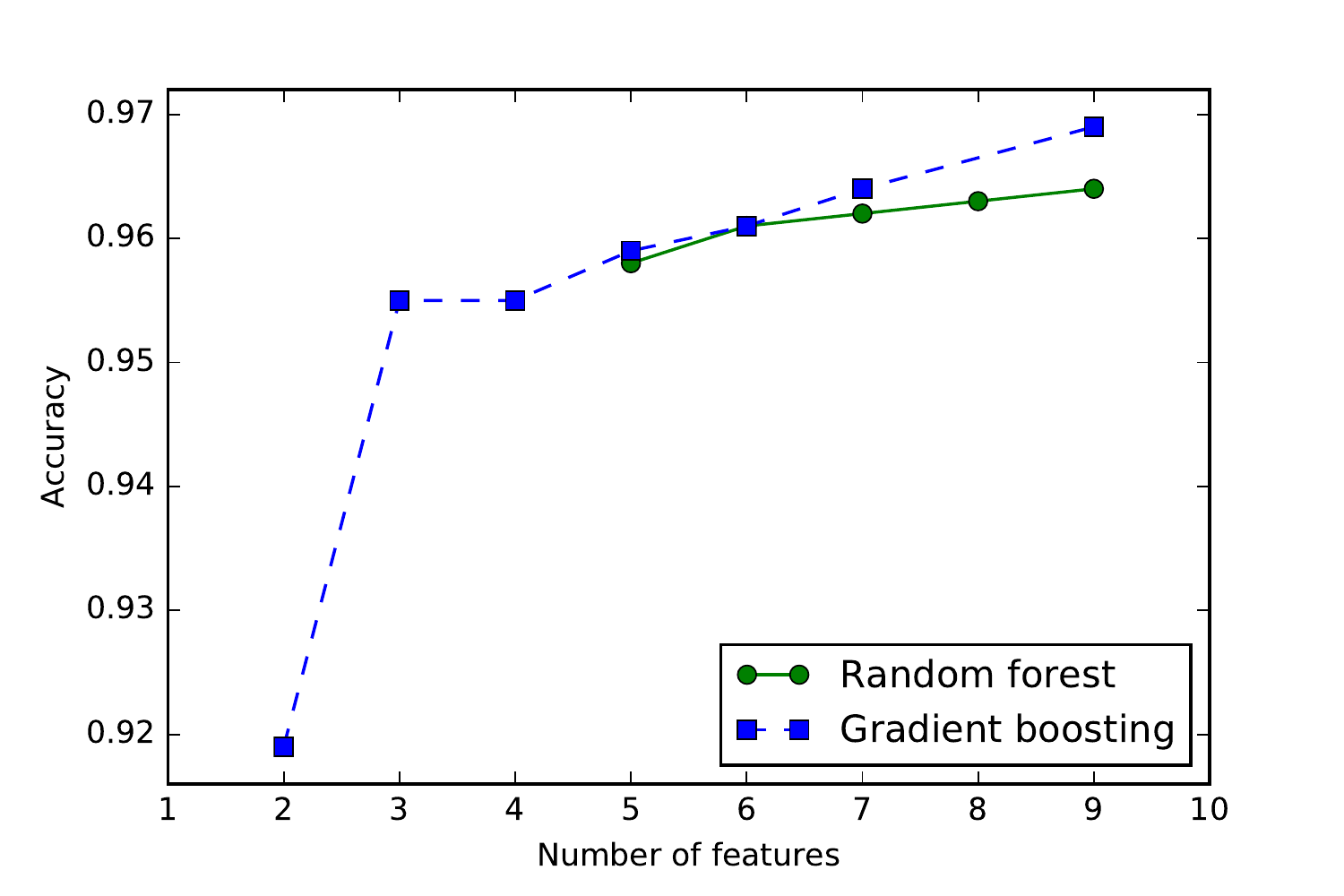}\        %accuracy_number_features.pdf
\caption{Comparison of the feature-based classifiers. (a) Number of features required to achieve  the given threshold of cumulative importance. (b) Accuracy of classifiers trained with the reduced number of features.\label{fig:fb comparison}}
\end{figure}

To elaborate on that issue, we first found the feature selection thresholds for cumulative importance levels ranging from 90\% up to 99\%. Then we trained both models again with the reduced number of features as indicated by the threshold. Results are presented in Fig.~\ref{fig:fb comparison}. As we see, gradient boosting reaches a given cumulative importance level with a smaller feature set than the random forest. Consequently, it requires less features to achieve high accuracies. 

\subsection{Deep-learning classification}

The \texttt{mcfly} package~\cite{KUP17} used in this work for the deep learning approach is a piece of software tailor-made to a classification of time series. One of its biggest advantages is a low entry level, because it does not require a user to define exactly the architecture of a convolutional neural network and to provide all hyper-parameters of the model. Instead, it carries out a search over suitable architectures and their hyper-parameters to find the best performing model. Since a diffusion trajectory is nothing but a multichannel time series (2D or 3D, depending on the problem at hand), it should match the requirements of \texttt{mcfly}.

\subsubsection{CNN architecture}

The first step in the development of a deep learning model is to create its architecture, i.e. to specify the following set of hyper-parameters: (i) the learning rate, (ii) the regularization rate, (iii) the number of convolutional layers, (iv) the number of filters in each convolutional layer and  (vi) the number of hidden nodes (in dense layers). The learning rate scales the magnitude of weight updates in the training process in order to minimize the network's loss function. The regularization helps to prevent overfitting of the network.

As it is not known a priori, which architecture will be optimal for classification of SPT trajectories,  we performed a random search to find the best one. This procedure simply creates a number of models at random, trains them on a relative small subset of the training data and then checks how good they are. Different criteria for selecting the best model are possible. The accuracy on a validation set will be used in this work.

Once the best model is chosen, it should be trained again on the full training data set. One full pass of the data through the network in the process of training its weights is called an epoch. Usually, many epochs are required to achieve a combination of the weights that yields a good accuracy.

We used \texttt{mcfly}'s function \verb?find_best_architecture? to perform the random search in the hyper-parameter space. We had to specify only two input parameters: the number of architectures for the random search procedure and the number of epochs for training the final model. As for the number of architectures, we expect intuitively that the bigger it is, the better. This is simply due to the fact that more architectures cover a larger part of the hyper-parameter space. Similarly, more epochs should guarantee a better convergence of the weights to the combination which minimizes the network's loss. Unfortunately, increasing the values of each of these parameters leads to significantly longer computing times, because the evaluation of an additional model as well as an epoch of training of the final model are very time consuming processes. Therefore, the choice of the values is usually a trade-off between the targeted accuracy and the computation time.

\begin{figure}
\includegraphics[scale=0.5]{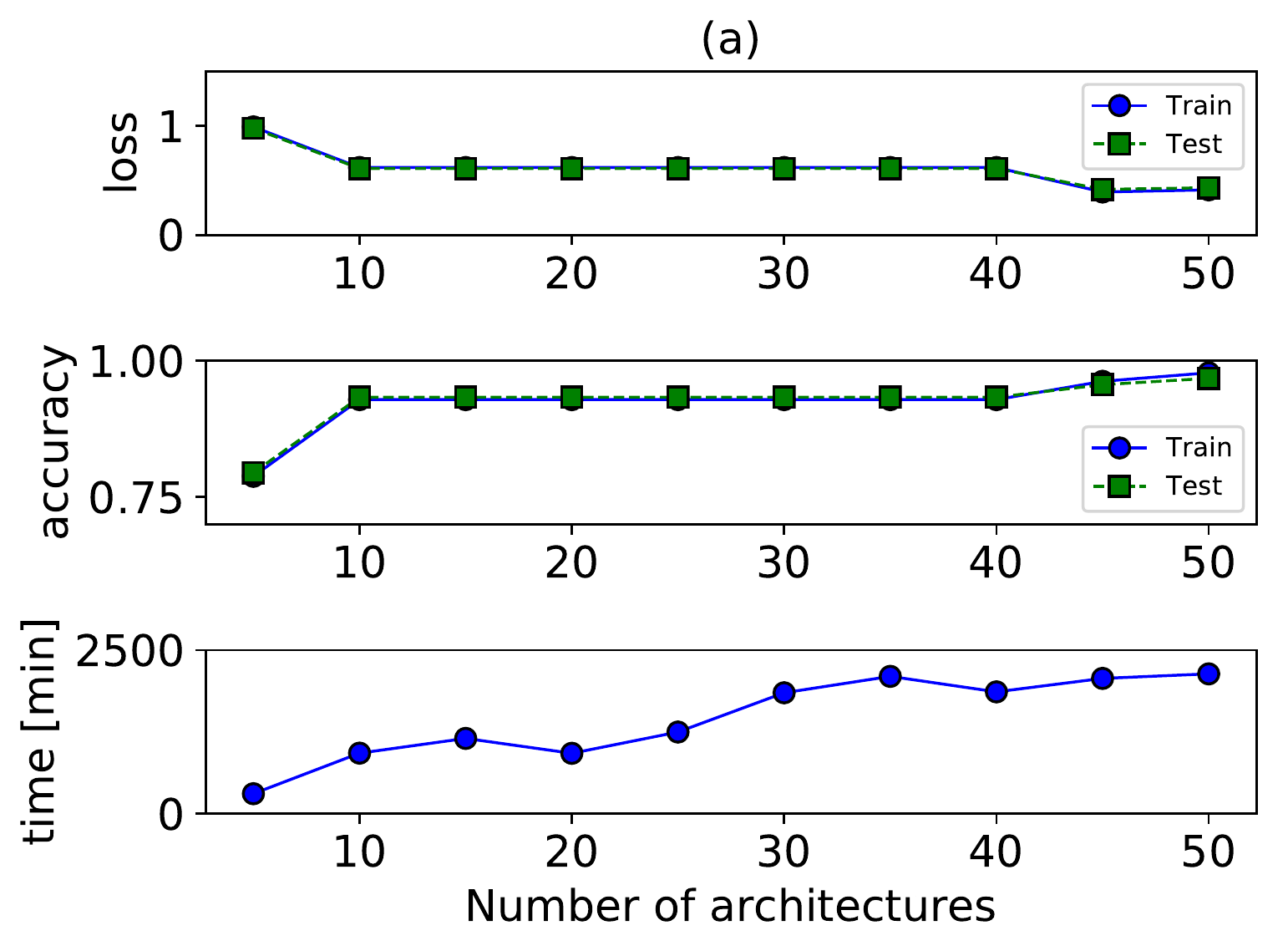}\ %search.pdf
\includegraphics[scale=0.5]{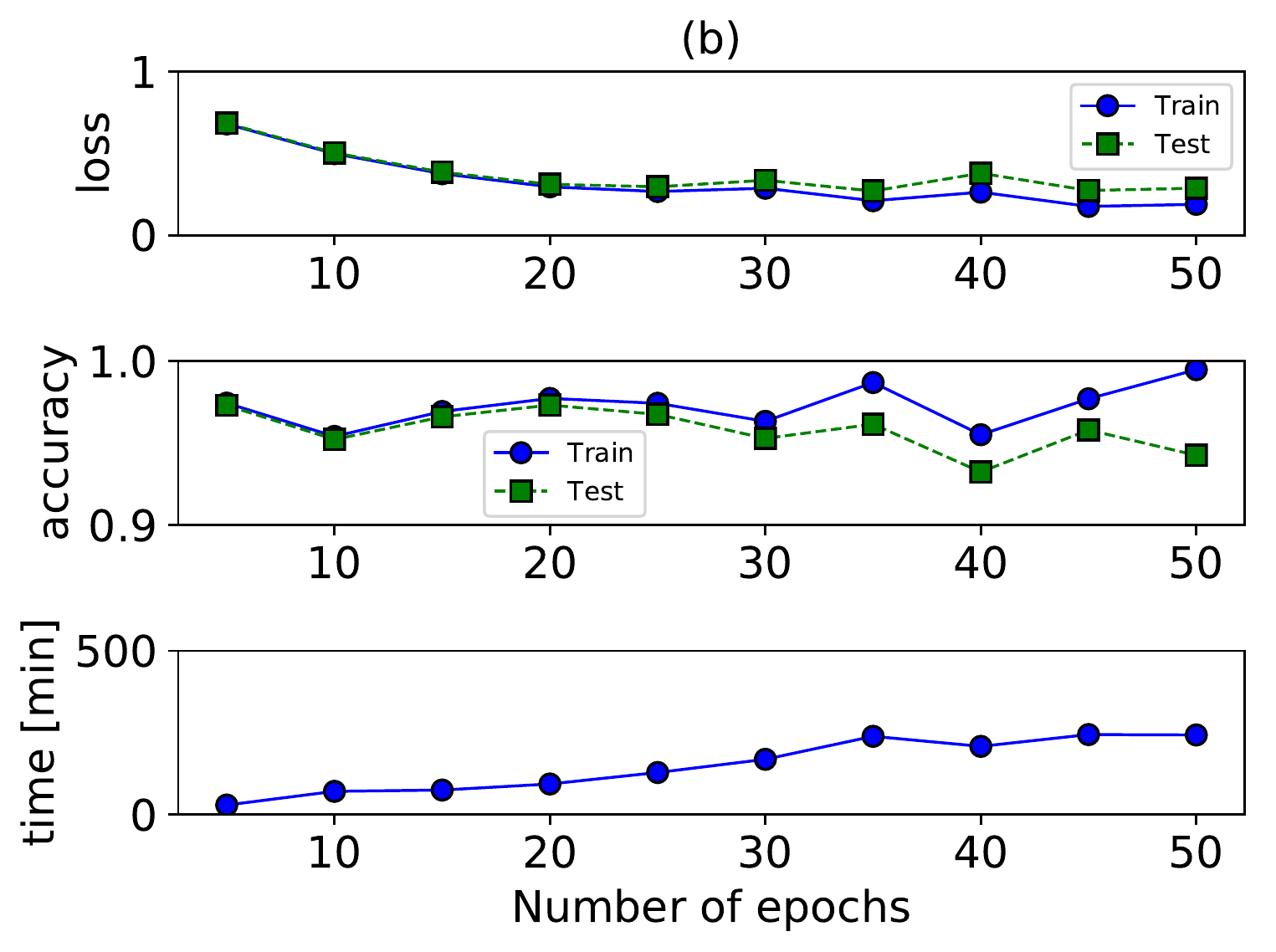} %epochs.pdf

\caption{(Color online) Impact of (a) the number of architectures in the random search and of (b) the number of epochs in training the final model on the loss, accuracy and execution time. In the random search procedure, 10 epochs were used to train the models on a small subset of the training data. The final model analyzed in the right column was selected from 20 architectures.\label{fig:search}}
\end{figure}

In order to determine the right values of the input parameters, we checked their impact on the loss and the accuracy of the final model as well as on the total execution time. Results are shown in 
Fig.~\ref{fig:search}. 
The analysis of the number of architectures (left column) was performed for 10 epochs on the training subset. As for the epochs (right column), the final model was selected from 20 initial architectures. First of all, we observe an almost monotonic growth of the execution time with the increase of both input parameters. As expected, the accuracy of the network increases with the number of architectures (middle left panel in Fig.~\ref{fig:search}). However, it remains practically constant for the values between 20 and 40 architectures. Thus, 20 architectures will be used in our further investigation, as it seems to be a good compromise between the accuracy and the execution time.

The behavior of the accuracy of the model as a function of the number of epoch (middle right panel in Fig.~\ref{fig:search}) is more interesting. We see that starting from 30 epochs, the difference between the accuracy on the train and the test datasets increases. This is an indicator that the model overfits, i.e. it starts to learn the noise in the training data as an important concept, which does not apply to the new trajectories from the test set. Since there is a local maximum in the accuracy at 20 epochs, we will use this value in the following.

\begin{figure}
\includegraphics[scale=0.4]{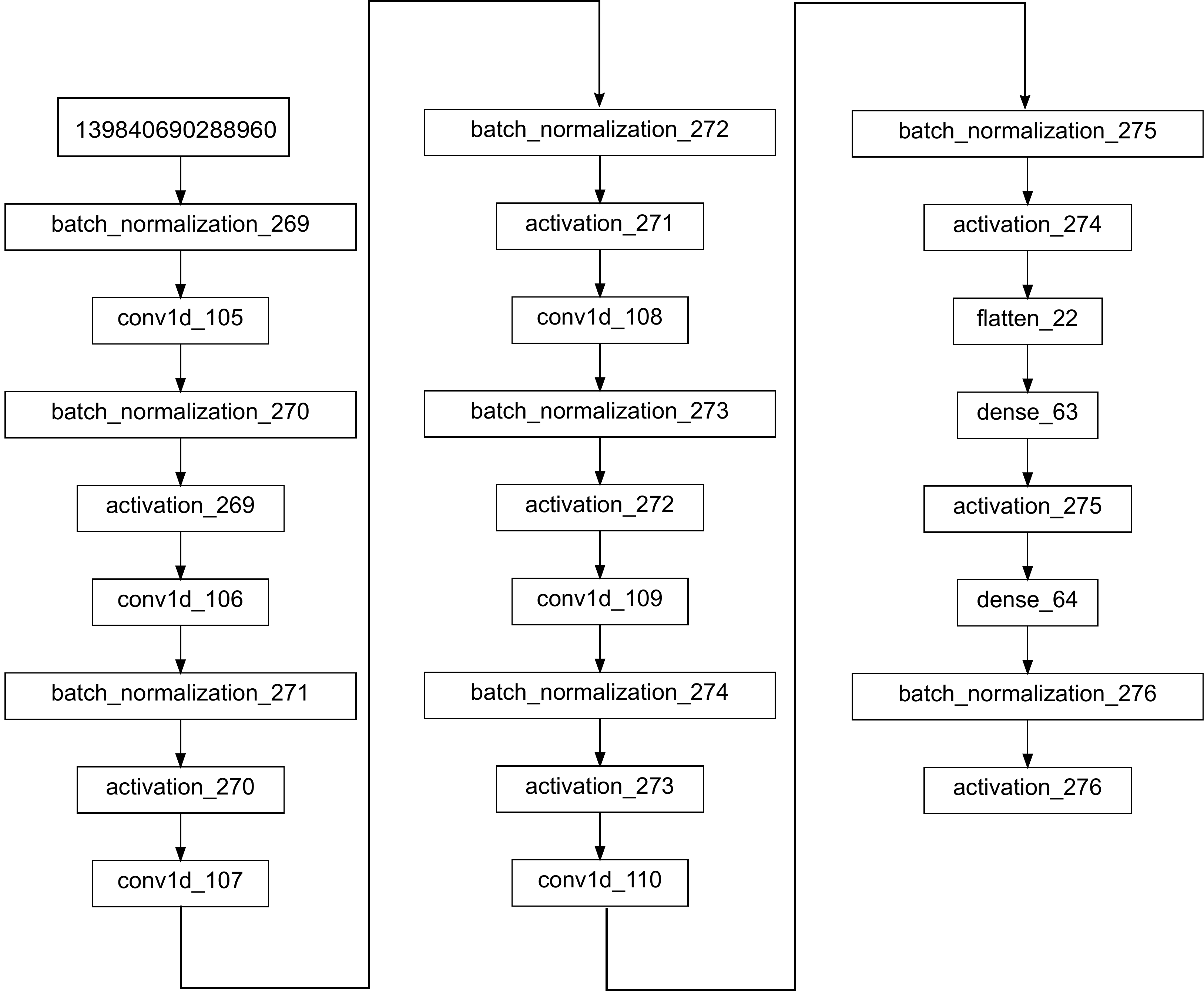} %architecture.pdf
\caption{Architecture of the best performing network model found by \texttt{mcfly}.\label{fig:cnn}}
\end{figure}

To summarize, our final model is the result of the random search among 20 architectures, trained for 20 epochs on the full training data set. Its architecture is shown in Fig.~\ref{fig:cnn}. It consist of 6 convolutional layers and 2 dense ones. Except those building blocks, there are several others elements of the model: (i) activation layers which define the output of neurons given an input or set of inputs, (ii) batch normalization layers responsible for normalization of the activation of previous layers, (iii) flatten layers, which flatten the input without changing its size (required by the dense layers). 
The values of the most important hyper-parameters of this model are summarized in Table~\ref{tab:cnn_params}.
\begin{table}
\begin{tabular}{l|c}
\hline\hline 
Parameter  & Values \\ 
\hline 
Regularization rate &  $0.0014064205292043147$ \\ 
\hline 
Number of $Conv$ layers & 6 \\
\hline
Number of filters &  $[49, 36, 18, 83, 90, 27]$ \\ 
\hline 
Learning rate &  $0.00021795428728036654$ \\ 
\hline 
Hidden nodes in dense layers & 1550\\
\hline\hline
\end{tabular} 
\caption{Hyper-parameters of the best performing network model shown in Fig.~\ref{fig:cnn}.\label{tab:cnn_params}}
\end{table}

\subsubsection{Accuracy of CNN}

\begin{figure}
\includegraphics[scale=0.5]{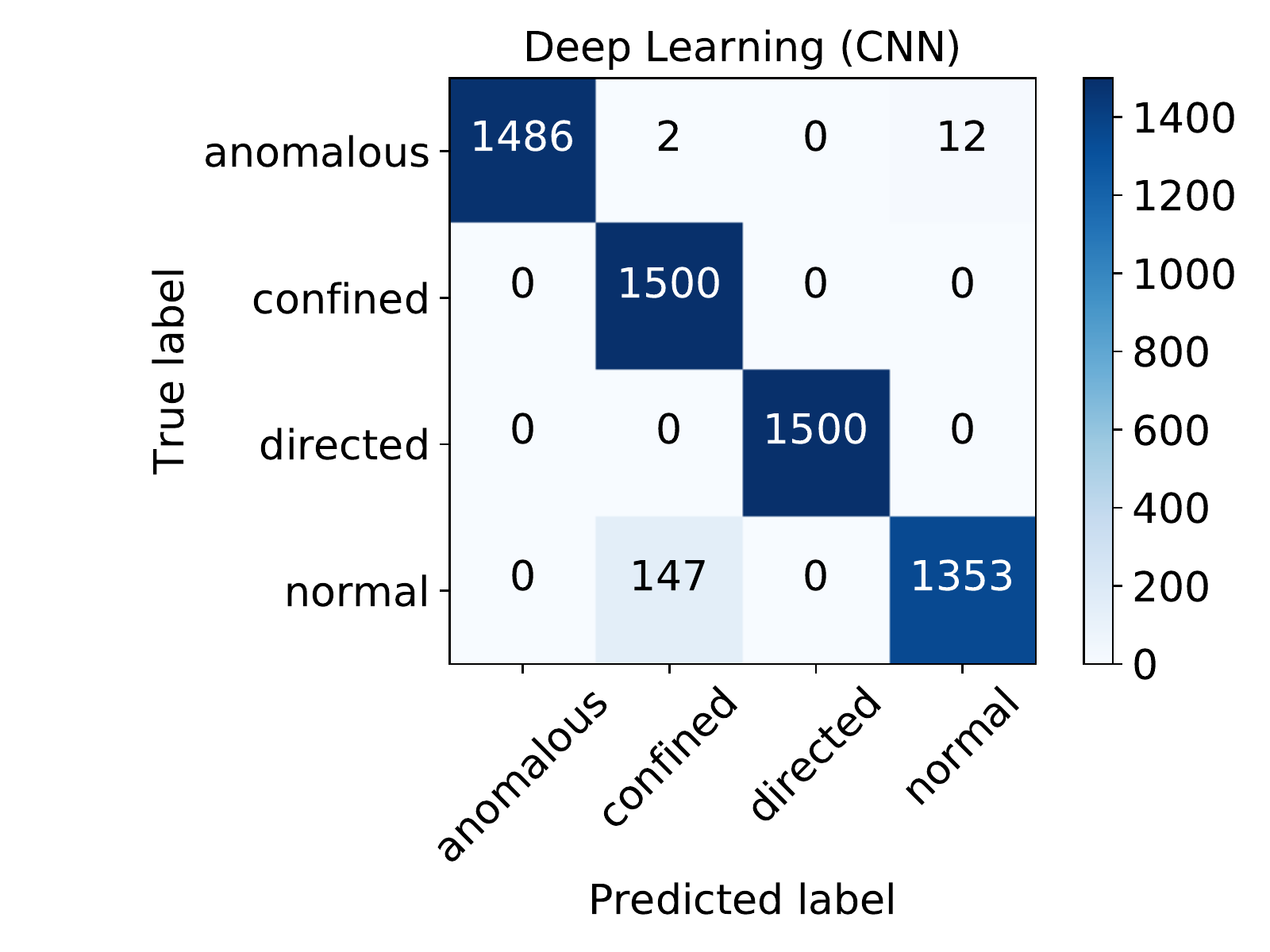}%cnn_cm.pdf
\caption{Confusion matrix of the CNN classifier.\label{fig:cm cnn}}
\end{figure}

The confusion matrix of our CNN model is shown in Fig.~\ref{fig:cm cnn} and its performance is summarized in Table~\ref{tab:report_dl}. 
\begin{table}
\begin{tabular}{c||c|c|c}
\hline \hline
 & precision & recall &  support  \\ 
\hline 
anomalous & 1.00 & 0.99 &  1500  \\ 
\hline 
confined & 0.91 & 1.00 &  1500 \\ 
\hline 
directed & 1.00 & 1.00 &  1500 \\ 
\hline 
normal & 0.99 & 0.90 &  1500 \\ 
\hline 
average/total & 0.98 & 0.97 &  6000 \\ 
\hline 
accuracy & \multicolumn{3}{c}{97.30\%}  \\ 
\hline \hline
\end{tabular} 
\caption{A brief summary of the performance of the CNN classifier. All results are rounded to two decimal digits.\label{tab:report_dl}}
\end{table}
Its overall accuracy turns out to be slightly better that the one of feature-based methods (see Table~\ref{tab:report_fb} for comparison). Again, the model not only returns much more relevant results than the irrelevant ones (high precision), but also yields most of the relevant results (high recall). The best performance is observed for the directed motion and the anomalous diffusion. It decays slightly for the confined diffusion in terms of precision and for the normal diffusion in terms of recall.

\subsection{Feature-based vs deep-learning}

Let us first juxtapose accuracies of the  methods analyzed in this paper together with the required processing times. All computations were carried out on a cluster of 24 CPUs (2.6 GHz each) with 50 GB of the total memory. Results are collected in Table~\ref{tab:comparison}. As already pointed out in the previous sections, the deep learning approach has a slightly higher accuracy than the feature based methods, however at costs of significantly longer processing times. Thus, if the time to train the model constitutes an issue, one should  rather go for gradient boosting. 
\begin{table}
\begin{tabular}{c||c|c||c}
\hline \hline
 & \multicolumn{2}{|c||}{Feature-based} & Deep-learning \\ 
\cline{2-4} 
 & Random forest & Gradient boosting & CNN \\ 
\hline 
Accuracy & $96.43\%$ & $96.97\%$ & $97.30\%$  \\ 
\hline 
Processing time & $1h26min$ & $1h9min$ & $3d5h50min$ \\ 
\hline \hline
\end{tabular} 
\caption{Comparison of all three classification methods. The processing time is understood as data preparation (if required), feature extraction (if required), searching for best performing model and finaly training and validation of the classifier. A cluster of 24 CPUs with 50GB total memory was used to perform the computations\label{tab:comparison}}
\end{table}

\begin{figure}
\includegraphics[scale=0.5]{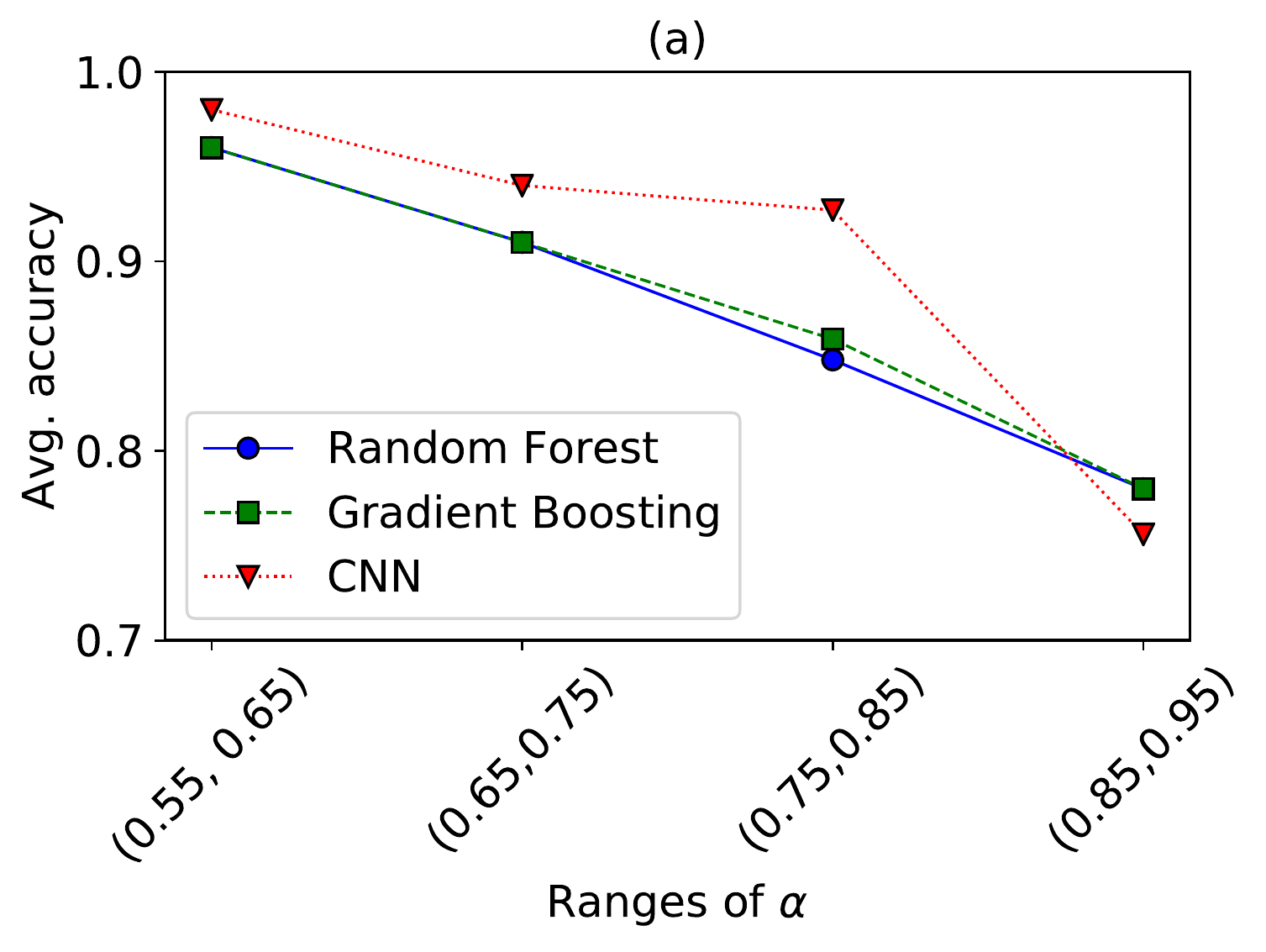}%anomalous.pdf\
\includegraphics[scale=0.5]{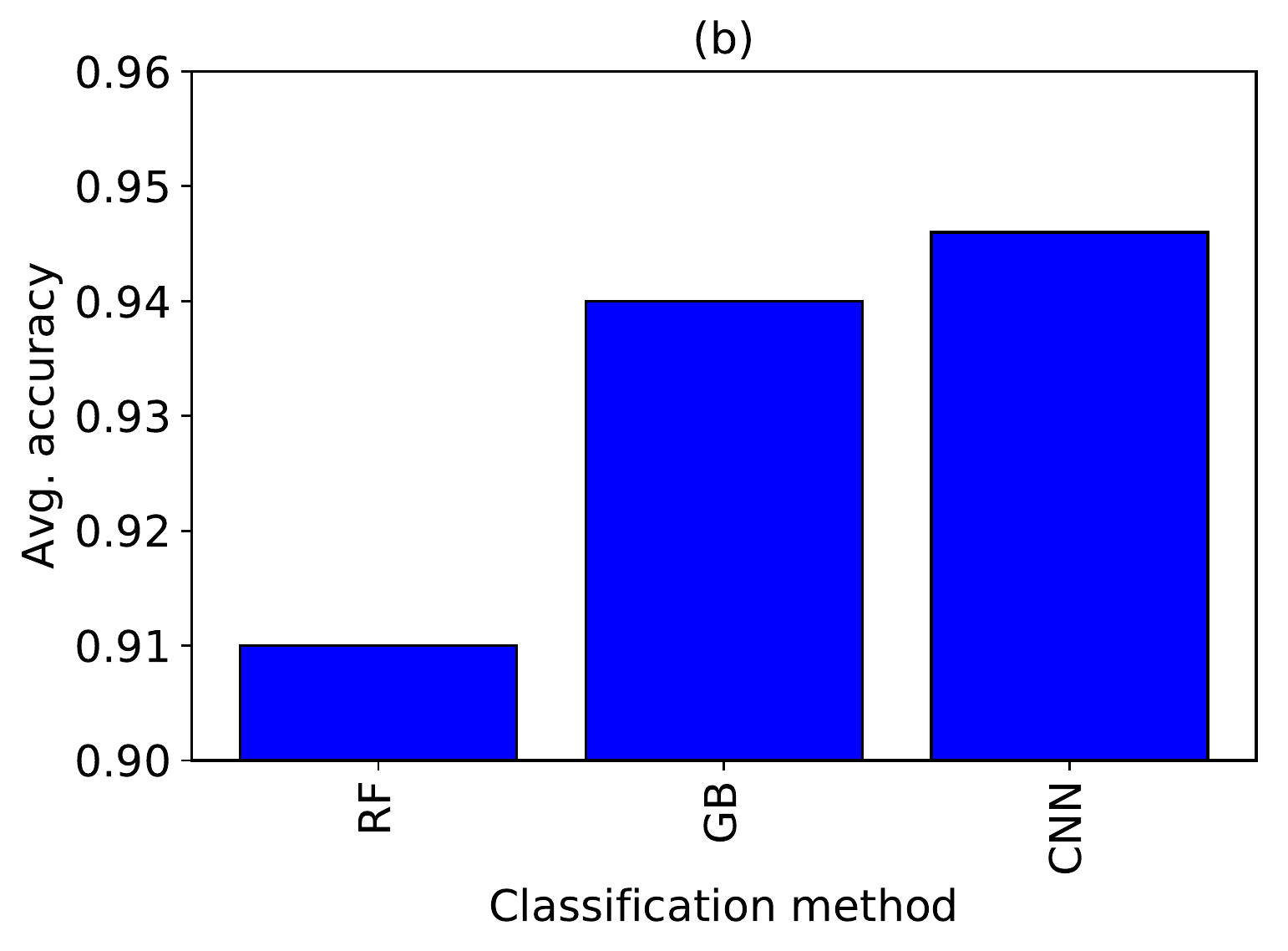}%directed.pdf\
\caption{Average accuracy of the methods in some limiting cases. (a) Performance of the classifiers for anomalous diffusion for four different ranges of the exponent $\alpha$. The lines in the plot are used to guide the eye. (b) Same for directed motion with small velocities, corresponding to $R\in\{1,2,3\}$ (see Eq.~(\ref{eq:r}) for the definition of $R$).   \label{fig:failure}}
\end{figure}

It would be interesting to see how the methods perform in some limiting cases, in which we expect them to fail anyway. For instance, an anomalous diffusion with the exponent $\alpha$ approaching 1 should be practically indistinguishable form a normal diffusion. Same holds for a directed motion with small velocities. To elaborate on that issue, we first generated four separate validation sets for the anomalous diffusion. Each set contained 1500 trajectories with the values of $\alpha$ randomly chosen from a corresponding interval: $\alpha^{(1)} \in (0.55,0.65\rangle $ for the first set, $\alpha^{(2)} \in ( 0.65,0.75\rangle $ for the second one, $\alpha^{(3)} \in ( 0.75,0.85\rangle $ for the third one and finally $\alpha^{(4)} \in ( 0.85,0.95\rangle$. Then we classified those sets with all three methods by making use of the already trained models. Results are shown in the left panel of Fig.~\ref{fig:failure}.
In case of the feature based methods we observe an almost linear decrease of the average accuracy with increasing $\alpha$. The CNN method performs better and the decrease is slower for  $\alpha < 0.85$. However, in the interval close to the limiting value ($\alpha=1$), there is a sudden drop in the performance of CNN and the deep learning approach starts to be the worst one.

We did a similar analysis for the directed motion with small velocities. This time, we generated only one additional validation set with $R\in \{1,2,3\}$ (see Eq.~(\ref{eq:r}) for the definition of $R$). Again, we classified it with all methods. Results are shown in the right panel of Fig.~\ref{fig:failure}. The CNN method turned out to be the best one, followed by the gradient boosting. Although the random forest still performs reasonably, there is already a noticeable gap in the accuracy to the other methods.

In this paper, we used four basic models of diffusion to generate artificial training data. However, they are not exhaustive and other models are possible for a given type of diffusion. For instance, FBM with $\alpha <1$ is not the only model that produces subdiffusive trajectories. Continuous time random walks (CTRW) with heavy-tailed waiting times~\cite{MAG09} or fractional Levy stable motion~\cite{BUR10} are known to have the characteristics of subdiffusion. Similarly, FBM with $\alpha >1$~\cite{BUR10} or CTRW with long-tailed spatial distribution~\cite{MAG12} are, alongside the directed motion model, examples of a superdiffusive process.

Now, one may for instance ask the question if a classifier trained on the directed motion model as the only expression of superdiffusion will recognize trajectories generated with other superdiffusive models. To check that, we took FBM with $\alpha \in (1.3,1.7)$ to generate an additional validation set consisting of 5000 trajectories. Results of their classification with our models are shown in Table~\ref{tab:superdiffusion}.
\begin{table}
\begin{tabular}{c||c|c|c|c}
\hline 
 & Anomalous & Confined & Directed & Normal \\ 
\hline \hline
Random forest & 206 & 399 & 1231 & 3164 \\ 
\hline 
Gradient boosting & 545 & 784 & 1380 & 2291 \\ 
\hline 
CNN & 0 & 33 & 9 & 4958 \\ 
\hline 
\end{tabular} 
\caption{Classification results for superdiffusive trajectories generated with a model other than the directed motion used to train the classifiers. In this particular example, FBM with $\alpha \in (1.3,1.7)$ was used to prepare the validation set. \label{tab:superdiffusion}}
\end{table}
We see that all methods perform really poor in this test. Only $28\%$ of the trajectories are classified as the directed motion (i.e. superdiffusion) by the gradient boosting, $25\%$ by the random forest and only $0.2\%$ by CNN. Thus, the models do not generalize well to unseen models, even though they are supposed to produce the same diffusion type as the ones used for training. This constitutes an issue in applications to real SPT data. Since it is rather impossible to provide a large set of experimental trajectories already labeled with correct diffusion types, we have to resort to artificially created training sets. As already shown, the machine learning methods work excellent on unseen and noisy data, but are not able to generalize well to unseen models. Therefore, we should include as many models as possible in our training sets in order to get some conclusive results for real trajectories.

\section{Conclusions}
\label{sec:conc}

We proposed a novel approach to analysis of SPT trajectories that makes use of convolutional neural networks, i.e. one of the popular modern deep-learning methods. The biggest advantage of this approach is that it works with raw SPT data. It does not require any complex data preprocessing nor the extraction of human-engineered features from data in order to feed a classifier. Instead, it learns the features on its own from the trajectories.

Deep-learning is seen already as the state-of-the-art classification method in many areas. From our results it follows that indeed, it has a slightly better accuracy than the traditional feature-based methods in most of the cases, but at costs of significantly longer training times.

We have shown that more models of diffusive processes have to be taken into account before applying ML to classification of trajectories. All methods considered in this paper perform excellent on unseen data, provided it was generated with the models already used in the preparation of the training sets. Unfortunately, they fail to correct classify trajectories produced with other models. Interestingly, CNN turned out to be the worst in this respect. Therefore, exhaustive data sets including as many models of diffusion as possible are needed to get conclusive classification results for real trajectories.

The excellent performance of the traditional methods observed in our experiments may be related to the fact that we assumed the movement of the particles to be homogeneous, i.e. one generated trajectory represents  only one type of motion. In real experiments the type of the diffusion may change multiple times within one trajectory due to the interaction of the particle with the medium. To cope with that issue one usually divides the trajectory in short segments and then tries to classify each segment independently of the others. Classifiers trained on data with short lengths are required for that purpose. The CNN method could work better than the feature based ones in this case, because most of the features relate to MSD estimates which are worse (much noisier) for short trajectories.

\begin{acknowledgments}
H. Loch-Olszewska and J. Szwabiński were supported by NCN-DFG Beethoven Grant No. 2016/23/G/ST1/04083. Computations were performed on the BEM cluster in the Wrocław Center for Networking and Supercomputing (WCSS).  
\end{acknowledgments}

\bibliography{spt_ml}

\end{document}